%% file: asahina_v1.tex
\documentclass[preprint]{aastex631}
\usepackage{graphicx,color,bm}
\usepackage[T1]{fontenc}
\definecolor{green}{rgb}{0,.6,0}

\def\mr#1{\mathrm{#1}}

\newcounter{graph}
\def\tr#1{#1}

\def\tc#1{#1}
\def\tg#1{#1}
\def\mr#1{\mathrm{#1}}

\def\tra#1{\textcolor{black}{#1}}

\def\tca#1{\textcolor{black}{#1}}
\def\trb#1{\textcolor{black}{#1}}

\begin{document}
\title{General relativistic radiation-MHD simulations of Precessing Tilted Super-Eddington Disks}
\author{Yuta \textsc{Asahina}$^{1}$ and Ken \textsc{Ohsuga}}

\email{asahinyt@ccs.tsukuba.ac.jp}

\affil{University of Tsukuba, 1-1-1 Tennodai, Tsukuba, Ibaraki 305-8577, Japan}

\input abstract_v1.tex

\input introduction_v1.tex
\input numerical_method_v1.tex
\input numerical_results_v2.tex
\input summary_v1.tex

\bigskip
\bibliography{apj-jour,apj4}
\newpage
\input figures_v1.tex

\end{document}

%% file: abstract_v1.tex
\begin{abstract}
We perform a three-dimensional general relativistic radiation magnetohydrodynamics simulation of a tilted super-Eddington accretion disk around the spinning black hole (BH).
The disk, that tilts and twists as it approaches the BH, 
precesses while maintaining its shape.
The gas is mainly ejected around the rotation axis of 
the outer part of the disk rather than around the spin axis of the BH.
The disk precession changes the ejection direction of the gas with time.
The radiation energy is also released in approximately the same direction as the outflow, so the precession is expected to cause a quasi-periodic time-variation of the observed luminosity.
The timescale of the precession is about $10\ \mr{s}$ for the 10 solar mass BH and for the radial extent of the disk of several tens of gravitational radii.
This timescale is consistent with
the frequency of the low-frequency quasi-periodic oscillation ($0.01-1\ \mr{Hz}$) observed in some ultraluminous X-ray sources.  
\end{abstract}

%% file: introduction_v1.tex
\section{Introduction}
The accretion disk forms around the black holes (BH) when the rotating gas accretes to a compact object such as X-ray binaries or active galactic nuclei. 
The gravitational energy of the accreting matter is released, and then a part of the released energy is converted to the thermal, magnetic, and radiation energy. 
As a result, it is thought that the strong radiation and jet appear.
To research the structure and dynamics of the accretion disk, magnetohydrodynamics simulations \citep{2002ApJ...566..164H,2008PASJ...60..613M}, radiation hydrodynamics simulations \citep{1988ApJ...330..142E,1997PASJ...49..679O,2005ApJ...628..368O}, and radiation magnetohydrodynamics simulations \citep{2009PASJ...61L...7O,2010PASJ...62L..43T,2011ApJ...736....2O} have been performed. 
Thereafter the effect of the general relativity was included \citep{2014MNRAS.441.3177M,2014MNRAS.439..503S,2016ApJ...826...23T,2016MNRAS.456.3929S}. 
These simulations assume that the rotation axis of the accretion disk aligns with the spin axis of the BH. 
However, the rotation axis would be tilted with respect to the BH spin axis, if the spin axis is not perpendicular to the orbital plane in the BH binary, for example. 
Furthermore, the BH spin and disk rotation axes may also be misaligned if gas accretes from a random direction onto an isolated BH. 
It is no guarantee that the spin axis of the supermassive BH in the galactic center aligns with the rotation axis of the accretion disk which is formed by the galaxy merger or galaxy-galaxy interaction. 

When the BH spin axis is not aligned with the rotation axis of the accretion disk, it has been pointed out that the frame-dragging effect can cause the precession of the accretion disk around the BH \citep[Lense-Thirring effect;][]{1975ApJ...195L..65B,1999ApJ...525..909A}. 
In fact, \tr{the disk precession has been reproduced by general relativistic hydrodynamics simulations \citep{2005ApJ...623..347F} and general relativistic magnetohydrodynamics simulations \citep{2007ApJ...668..417F}.}
\tr{
These simulations have shown that the disk precession might be responsible for the low-frequency quasi-periodic oscillations \citep[e.g.][]{1999ApJ...524L..63S}.
\cite{2018MNRAS.474L..81L} revealed that the jet which is powered by Blandford-Znajek \citep{1977MNRAS.179..433B} mechanism precess with the disk precession. 
This result implies that the Lense-Thirring precession might be the origin of the wiggling jet such as that observed in the radio galaxy 3C31 \citep{2008MNRAS.386..657L}.
}
Although these studies \tr{treat} the geometrically thick disk, general relativistic magnetohydrodynamics simulations of the geometrically thin disk \tc{have shown} that several sub-disks are formed by disk tearing due to the Lense-Thirring effect \citep{,2019MNRAS.487..550L,2021MNRAS.507..983L,2023MNRAS.518.1656M}. 

The phenomena which can originate from the disk precession are also observed in the super-Eddington sources although previous studies mentioned above assume the low mass accretion rate. 
One of these phenomena is the rapidly-changing jet orientation observed in V404 Cygni whose luminosity is thought to be higher than the Eddington luminosity. 
The propagation direction of this jet changes rapidly in a few minutes or hours \citep{2019Natur.569..374M}. 
In addition, luminosity oscillations with $0.01-1\ \mr{Hz}$ observed in the ultraluminous X-ray sources (ULXs) \citep{2019MNRAS.486.2766A} can be explained by the precession of the accretion disk,
although obscuration by the clumpy clouds in the radiatively-driven disk winds passing across our line of sight is also a possibility \citep{2011MNRAS.411..644M,2013PASJ...65...88T,2018PASJ...70...22K}.
We need to perform general relativistic radiation magnetohydrodynamics (GR-RMHD) simulations to study the precession of the luminous accretion disks.  
Recently, GR-RMHD simulations of the tilted accretion disk have been performed by \cite{2023ApJ...944L..48L}.
However, the accretion rate is lower than the Eddington rate ($L_\mr{Edd}/c^2$) in their simulations,
where $L_\mr{Edd}$ is the Eddington luminosity,
so that the precession of tilted super-Eddington disk still has not been studied. 

We perform a GR-RMHD simulation of the tilted super-Eddington accretion disk with the mass accretion rate of about \tr{$300L_\mr{Edd}$}. 
We introduce the basic equations and initial and boundary conditions in Section 2. 
The numerical results 
are shown in Section 3. 
Finally, Section 4 is devoted to a summary and discussion. 

%% file: numerical_method_v1.tex
\section{Numerical Method}
\label{Numerical Method}
\subsection{Basic Equations}
\label{Basic Equations}
In this paper, we numerically solve GR-RMHD equations using the code developed by \cite{2016ApJ...826...23T}. 
We hereafter take the light speed $(c)$ and gravitational constant $(G)$ as unity. 
The Greek and Latin suffixes indicate spacetime and space components, respectively. 
Mass conservation equation, energy-momentum conservation equation for magnetofluids, 
induction equation, and energy-momentum conservation equation for radiation are as follows\tr{:} 
\begin{equation}
  (\rho u^{\nu})_{;\nu} = 0\tr{,} 
  \label{eq.cons_mass}
\end{equation}
\begin{equation}
  {T^{\nu}}_{\mu;\nu} = G_{\tr{\mu}}\tr{,} 
  \label{eq.cons_EM_MHD}
\end{equation}
\begin{equation}
  (b^{\nu}u^{j}-b^{j}u^{\nu})_{;\nu} = 0\tr{,} 
  \label{eq.Bmag}
\end{equation}
\tr{and}
\begin{equation}
  {R^{\nu}}_{\mu;\nu} = - G_{\mu}\tr{.} 
   \label{eq.cons_EM_Rad}
\end{equation}
$\rho$, ${R^{\nu}}_{\mu}$, $u^{\mu}$, and $b^{\mu}$ are the gas density, the energy-momentum tensor for radiation, four-velocity, and the covariant magnetic field, respectively.
We employ the Kerr metric with the black hole spin ($a$) of 0.9 in the Kerr-Schild coordinate ($r$, $\theta$, $\phi$). 
The mass of the black hole $(M)$ which is assumed to be $10M_{\odot}$, where $M_{\odot}$ is the solar mass.  
${T^{\mu}}_{\nu}$ is the energy-momentum tensor for magnetofluids,
\begin{equation}
  {T^{\mu}}_{\nu} = \left( \rho + \frac{\Gamma}{\Gamma-1}p + \frac{b^{2}}{4\pi}  \right) u^{\mu}u_{\nu} - \frac{b^{\mu}b_{\nu}}{4\pi} + \left( p + \frac{b^{2}}{8\pi}  \right)\delta^{\mu}_{\nu} .
  \label{EM_MHD}
\end{equation}
$p$ is the gas pressure, $\delta^{\mu}_{\nu}$ is the Kronecker delta, and $\Gamma$ is the specific heat ratio which is set to be 5/3. 
$G_{\nu}$ is radiation four-force, 
\begin{equation}
  G_{\nu} = -\rho \kappa_{\mr{abs}} \left( R_{\nu\alpha}u^{\alpha} + 4\pi B u_{\nu} \right) - \rho\kappa_{\mr{sca}} \left( R_{\nu\alpha}u^{\alpha} + R_{\alpha\beta}u^{\alpha}u^{\beta}u_{\nu} \right).  
  \label{eq.Gmunu}
\end{equation}
Here, the free-free emission/absorption and isotropic electron scattering are considered.
The absorption and scattering opacities in the comoving frame are
$\kappa_{\mr{abs}} = 6.4 \times 10^{22}\rho T_{\mr{gas}}^{-7/2}\ \mr{cm^{2}g^{-1}}$
and
$\kappa_{\mr{sca}} = 0.4\ \mr{cm^{2}g^{-1}}$, where $T_\mr{gas}$ is the gas temperature obtained by the equation of state $p=\rho k_{\mr{B}} T_{\mr{gas}}/(\mu m_{\mr{p}})$. 
Here, $k_\mr{B}$ is the Boltzmann constant, $m_{\mr{p}}$ is \tr{the proton mass}, and $\mu$ is a mean molecular weight which is set to be 0.5.
The blackbody intensity ($B$) is given by 
\begin{equation}
  B = \frac{a_{\mr{rad}}T_{\mr{gas}}^{4}}{4\pi}, 
\end{equation}
where $a_{\mr{rad}}$ is the radiation constant. 
To close the equation \ref{eq.cons_EM_Rad} we adopt the M-1 closure scheme \citep{2007A&A...464..429G}. 
The radiation energy momentum tensor is given by 
\begin{equation}
  R^{\mu\nu} = 4p_\mr{rad}u^{\mu}_\mr{rad}u^{\nu}_\mr{rad}+p_\mr{rad}g^{\mu\nu}, 
\end{equation}
where $p_\mr{rad}$ and $u^{\mu}_\mr{rad}$ are the radiation pressure and radiation frame's four-velocity, respectively. 

When calculating the change in energy-momentum of radiation and magnetofluids via the radiation four-force, we solve the entropy equation instead of the energy conservation equation \citep[see section 4.7 in][]{2020ApJ...901...96A}. 

\subsection{Initial and Boundary Conditions}
\label{Initial and Boundary Conditions}

We assume the equilibrium rotating torus \citep{1976ApJ...207..962F} \tr{with the polytropic index of $5/3$}. 
We set \tr{the inner and outer radii of the torus to be $20 r_\mr{g}$ and $80r_\mr{g}$}, 
where $r_\mr{g}$ is the gravitational radius ($r_\mr{g}=M$). 
The maximum density \tr{of the torus which is $\rho_{0}= 10^{-2}\ \mr{g\ cm^{-3}}$} locates at $r=33 r_\mr{g}$ on the equatorial plane. 
We tilt the torus with respect to the spin axis of the BH. 
The initial tilt angle is $\mathcal{T}_0=30^\circ$ and the precession angle is $\mathcal{P}_0=180^\circ$. 
\tr{Here, we set the azimuthal component of the vector potential to be $A_\mr{\phi} = \max(\rho/\rho_0 - 0.2, 0)$ and the other components are set to zero.
With this setting, the single-looped poloidal magnetic field, of which the minimum ratio of the gas pressure to the magnetic pressure is $100$, is embedded in the torus. }
We assume that the initial radiation energy density is very low and that the radiation is \tr{locally} isotropic in the zero angular momentum observer frame. 
\tr{In this setting, the thermal energy of the gas is converted to the radiation energy in the torus by the emission immediately after the simulation starts. The gas temperature then becomes approximately equal to the radiation temperature. Because the torus is optically so thick that photons can hardly escape, the total pressure changes very little. Therefore, the torus is not significantly deformed.}

The simulation region is $r_\mr{in} \leq r \leq r_\mr{out}$, $\theta_0 \leq \theta \leq \pi-\theta_0$, and $0\leq\phi<2\pi$. 
Here $r_\mr{in} = \left( 1+\sqrt{1-a^2}\right) r_\mr{g}$, $r_\mr{out}=\tr{10^3}\ r_\mr{g}$, and $\theta_0=\pi/90$. 
The number of the grid points is set to be $(N_{r}, N_{\theta},N_{\phi})=(\tr{250}, 180, 64)$ in the space. 
\tr{In order to increase the resolution in the area where accretion disks are formed, especially near the BH, the non-uniform grid is employed in the $r$- and $\theta$-directions. Specifically, we set} $r=r_\mr{in} (r_\mr{out}/r_\mr{in})^{x_{1}}$ and \tr{$\Delta\theta = AC-0.5A\tanh{(10x_2)} -0.5A\tanh{[10(1-x_2)]}$ with $A\sim1.76(\pi-2\theta_0) \Delta x_2$ and $C\sim1.5$. We derive constants $A$ and $C$ to satisfy the conditions of $\theta =\theta_0$ at $x_2=0$ and $\theta=\pi-\theta_0$ at $x_2=1$}. 
Here $x_{1}$ and $x_{2}$ are \tr{both set at uniform intervals between 0 and 1.
A uniform grid is employed in $\phi$-direction.}

The outflow (inflow) boundary condition is adopted at the outer (inner) boundary at $r=r_\mr{out} (r_\mr{in})$. 
Then, $u^r=\max(u^r,0) [u^r=\min(u^r, 0)]$ and the other physical quantities are set to have no gradient. 
\tra{Note that we adopt the first-order scheme in three grids just outside the inner boundary (event horizon). 
\tca{Furthermore, we use the upwind method when calculating the numerical flux at the inner boundary.
As a result, the flux at the event horizon is taken as the flux at the innermost grid, and no information inside the event horizon is required.}
}
Therefore, we do not prepare the ghost grids inside the inner boundary in the present simulation.
Also, the transmissive boundary condition is employed at $\theta=\theta_0$ and $\theta=\pi-\theta_0$.

%% file: numerical_results_v2.tex
\section{Numerical Results}
\subsection{Overall Structure of Tilted Super-Eddington Accretion Disk}

\tr{When we start the simulation,} the differential rotation of the torus makes the toroidal magnetic field. 
The angular momentum is transported radially outward by the growth of the magnetorotational instability (MRI), then the gas accretion occurs. 
As a result, the super-Eddington accretion flow with the mass accretion rate of about $\tr{300}L_\mr{Edd}$ forms in our simulation. 
The equatorial plane of the accretion disk matches that of the initial torus in the region of $r\gtrsim 15r_\mr{g}$. 
On the contrary, the accretion disk within $r\sim 15r_\mr{g}$ has complex structures (see below for details). 

The black line in figure \ref{t-mdot} shows the mass accretion rate $(\dot{M}_\mr{in})$ at $r=r_\mr{in}$, \tr{which} is defined as 
\begin{equation}
    \dot{M}_\mr{in}=-\int^{2\pi}_0\int^\pi_0 \rho u^r \sqrt{-g} d\theta d\phi. 
\end{equation}
We find that the accretion rate exceeds the Eddington rate, $\dot{M}_\mr{in}\sim \tr{300}L_\mr{Edd}$, and is almost constant after $t \sim \tr{10^4} t_\mr{g}$. 
The radiation energy swallowed by the BH per unit time (trapped luminosity) is evaluated as
\begin{equation}
    L_\mr{rad,BH}=\int^{2\pi}_0\int^\pi_0 R^r_t \sqrt{-g} d\theta d\phi,
\end{equation}
\tr{at $r=r_\mr{in}$}, 
and is shown by the dashed red line in figure \ref{t-mdot}. 
It is clearly understood that most of the radiation energy generated by the release of the gravitational energy is swallowed due to the photon trapping since the trapped luminosity is much larger than the photon luminosity at $r=\tr{800}r_\mr{g}$ ($\sim \tr{1.7}L_\mr{Edd}$),
\tr{where the photon luminosity} is evaluated by
\begin{equation}
    L_\mr{rad}=-\int^{2\pi}_0\int^\pi_0 R^r_t \sqrt{-g} d\theta d\phi.
\end{equation}
The photon trapping is a characteristic feature in the super-Eddington accretion disk \citep{2005ApJ...628..368O,2015PASJ...67...60T}. 
The green line shows the time evolution of the electromagnetic luminosity,
\begin{equation}
    L_\mr{mag}=-\frac{1}{4\pi}\int^{2\pi}_0\int^\pi_0 \left(  b^2 u^r u_t -b^r b_t \right) \sqrt{-g} d\theta d\phi,
\end{equation}
measured at $r=800r_{\rm g}$. 
This is smaller than $10^{-2} L_\mr{Edd}$ at $t \gtrsim 2.5\times 10^4 t_\mr{g}$. 
We define the kinetic luminosity as
\begin{equation}
    L_\mr{kin}=-\int^{2\pi}_0\int^\pi_0 \rho u^r \left( 
u_t+\sqrt{-g_{tt}} \right) \sqrt{-g} d\theta d\phi, 
\end{equation}
\citep{2016MNRAS.456.3915S}. 
We note that the kinetic luminosity here is obtained by subtracting from the energy-momentum tensor ${T^r}_t$ the components corresponding to the thermal energy, rest mass energy, magnetic energy, and potential energy.
The blue line shows the time evolution of $L_\mr{kin}$ at $r=800r_{\rm g}$. 
\tr{The kinetic luminosity once increases to $\sim L_\mr{Edd}$, but then decreases. 
After $t\sim 1.5\times 10^4 t_\mr{g}$, $L_\mr{kin}$ becomes almost constant at $0.5L_\mr{Edd}$. 
The kinetic luminosity with $Be\geq0.05$ (jet region) is about $0.4L_\mr{Edd}$ which is about $80\%$ of $L_\mr{kin}$. 
This means that the jet has most of the outward kinetic energy at $r=800r_\mr{g}$. 
$Be$ is the Bernoulli parameter,}
\begin{equation}
    Be=-\frac{\rho u^r +{T^r}_t+{R^r}_t}{\rho u^r}. 
\end{equation}
When $Be$ is 0.05, the velocity of the gas at infinity becomes $0.3c$. 

\tr{
\tra{Figure \ref{r-lum} shows the radial profile of the time-averaged luminosity between $3.8\times10^4t_\mr{g}$ and $4.0\times10^4t_\mr{g}$. }
The kinetic luminosity is almost independent of the radius in the outer region,
$r \gtrsim$ a few $\times 100r_{\rm g}$. 
We find it is about $0.5L_\mr{Edd}$ \tra{in $r\gtrsim300r_\mr{g}$}. 
This is because the jet travels almost straight outward in the radial direction without much acceleration or deceleration.
Here, we note that the photon luminosity is also insensitive to the radius. It is \tra{$\sim L_\mr{Edd}$ in $r\gtrsim200r_\mr{g}$.}
The radiation flux is mildly collimated and the radiation energy is mainly released at the region of $Be\geq0.05$.
Indeed we find at $r=800r_\mr{g}$ that the photon luminosity for $Be\geq0.05$ is about $\sim 1.1 L_\mr{Edd}$, which is about 70\% of $L_{\rm rad}$.
Since the gas-radiation interaction is not very effective due to the small optical depth of the region of $Be>0.05$ (typically $\sim 0.14$), the photon luminosity is kept approximately constant \tra{(see Figure \ref{ur-ro})}.
}

\tr{
The energy conversion efficiency of the flow defined as 
\begin{equation}
\eta = 1 - \frac{\int^{2\pi}_{0}\int^{\pi}_{0} ({T^r}_t+{R^r}_t)\sqrt{-g}d\theta d\phi}{\dot{M}_\mr{in}}
\end{equation}
is about 0.54\% at $r=800r_{\rm g}$. 
The system releases energy primarily as radiation or jets.
Indeed, at $r=800r_{\rm g}$,
the outward energy efficiency via the radiation,
\begin{equation}
    \eta_{\rm rad}=-\frac{\int^{2\pi}_{0}\int^{\pi}_{0} {R^r}_t\sqrt{-g}d\theta d\phi}{\dot{M}_\mr{in}}
\end{equation}
is $0.42$\%, that by the mass outflow is 
\begin{equation}
    \eta_{\rm kin}=-\frac{\int^{2\pi}_{0}\int^{\pi}_{0} \rho u^r \left(u_t + \sqrt{-g_{tt}} \right)\sqrt{-g}d\theta d\phi}{\dot{M}_\mr{in}}
\end{equation}
is $0.12$\%, and their sum is approximately equal to $\eta$.
It is also found that the energy is mainly released through the jet region ($Be\geq0.05)$
since the outward energy efficiency via the radiation and outflow in that region are 
0.29 \% and 0.10 \%.
Here we note that the energy released at the event horizon is not reaching far enough and accumulates along the way.
This is evident from the fact that $\eta$ at the event horizon is $\sim 7$\%, much larger than at $r=800r_{\rm g}$.
We also note that $\eta\sim 7\%$ is roughly comparable with $\sim10\%$ for weak magnetic field model in \cite{2018MNRAS.474L..81L} and $\sim 15\%$ in \cite{2023ApJ...944L..48L}.
Since $\eta_{\rm rad}+\eta_{\rm kin}$ is $3.7$\% at $r=100r_{\rm g}$, about half of the energy released at the horizon accumulates in regions of $r<100r_{\rm g}$, and the other half in regions between $100r_{\rm g}$ and $800r_{\rm g}$. 
We can estimate the accumulated energy in the computational domain between $3.9\times 10^4t_\mr{g}$ and $4\times 10^4t_\mr{g}$ is about $1.3\times 10^{39} \mr{erg}$. 
In fact, we find that the energy with $Be<0$ in $2r_\mr{g}<r<10^3 r_\mr{g}$ increases with $1.1\times 10^{39} \mr{erg}$. 
This means that most of the accumulated energy is transferred to the bounded gas. 
\trb{
However, such energy accumulation is transient. The total energy in the simulation box does not change significantly, only slightly increasing or decreasing. In fact, it is about $1.61\times10^{41}$ erg at $t=10^4t_\mr{g}$, $1.57\times10^{41}$ erg at $t=2\times10^4t_\mr{g}$, $1.58\times10^{41}$ erg at $t=3\times10^4t_\mr{g}$, and $1.61\times10^{41}$ erg at $t=4\times10^4t_\mr{g}$. These fluctuations may be due to the fact that the outflow has not reached a steady state. Long-term simulations are needed for detailed verification.
}
}

Figure \ref{volume-rendering} shows the volume rendered image at $t=\tr{1.5\times10^4t_\mr{g}}$. 
Blue and orange show the density and Lorentz factor  $(v>0.3c)$, respectively. 
The direction of the BH spin is aligned with the vertical direction. 
We find the high density tilted accretion disk $(\rho\sim\rho_0)$ colored by blue and also the \tr{jet} (orange) that is blown away in a direction closer to the rotation axis of the accretion disk than to the BH spin axis (left panel).
The right panel of Figure \ref{volume-rendering} shows the density near the BH presented by a white filled circle.
The accretion disk has the non-axisymmetric structure. 
Especially the two-armed structure appears in the vicinity of the BH. 
Such structure has been confirmed in GR-MHD simulations, and its origin is thought to be the $\theta$ dependence of the innermost stable circular orbit (ISCO) radius \citep{2007ApJ...668..417F}. 
The ISCO radius increases as it approaches the BH spin axis. 
Thus, the disk matter first reaches the ISCO radius at $\phi=0^\circ, 180^\circ$, from where it flows into the BH. 
As a result, the two-armed accretion structure forms. 
In our simulation, disk tearing shown in \cite{2019MNRAS.487..550L} does not occur similarly to \cite{2007ApJ...668..417F}. 
This is probably because the accretion disk is geometrically thick. 

\tra{Figure \ref{ur-ro} shows the profiles of radial velocity at $\phi=270^\circ$ and density at $\phi=90^\circ$. 
\tca{The rotation axis of the disk lies in this plane and is tilted with respect to the BH spin axis.}
The jet propagates align with the rotation axis of the disk as shown in left panel. 
The white contour shows the photosphere measured from the outer boundary. 
The radius of the photosphere is about $100-200r_\mr{g}$ in the jet region and is about $800r_\mr{g}$ in the other region. 
The photon luminosity becomes independent on radius in $r\gtrsim 200r_\mr{g}$ shown as Figure \ref{r-lum} since the jet region is optically thin \tca{and radiation propagates freely}. 
\trb{We estimate the position of the photosphere by extrapolating the radial profile of the density when we perform simulations with a larger computational domain. 
We derive the distance of the photosphere by integrating the optical depth from infinity. 
The distances are roughly $3\times10^2 r_\mathrm{g}$ for the jet region and $10^3 r_\mathrm{g}$ for the other region. 
}
}

Figure \ref{r-tilt,prec} shows the radial profiles of the time-averaged tilt angle $(\langle \mathcal{T} \rangle)$ and precession angle $(\langle \mathcal{P} \rangle)$ between \tr{$3.8\times 10^4 t_\mr{g}$ and $4.0\times10^4 t_\mr{g}$} to understand the overall structure of the accretion disk. 
We evaluate tilt angle $(\mathcal{T})$ and precession angle $(\mathcal{P})$ as 
\begin{equation}
\mathcal{T}=\arccos{\left(\frac{L_z}{L}\right)}
\end{equation}
and
\begin{equation}
\mathcal{P}=\arctan{\left(\frac{L_x}{L_y}\right)},
\end{equation}
respectively. 
Here 
\begin{equation}
{\bm L}=(L_x,L_y,L_z)=\int^{2\pi}_0\int^\pi_0 \rho{\bm r}\times{\bm v}\sqrt{-g} d\theta d\phi
\end{equation}
is the angular momentum vector and ${\bm v}$ is the velocity in the observer rest frame. 
The shaded region indicates $1\sigma$-variation.  

The tilt angle is about \tr{$25^\circ$} in $r\gtrsim 15r_\mr{g}$, which \tr{roughly} matches the tilt angle of the initial torus $\mathcal{T}_0=30^\circ$. 
On the contrary, the precession angle is about \tg{$296^\circ$}, which is \tg{$116^\circ$} larger than the initial precession angle $\mathcal{P}_0=180^\circ$. 
This discrepancy is caused by the Lense-Thirring precession (we will describe details of the time evolution below.) 
In the regime of $5r_\mr{g}\lesssim r\lesssim 15r_\mr{g}$, 
the closer to the BH, the larger the tilt angle. 
This is induced by the fact that the gas accretes from the higher latitude region since the ISCO radius becomes larger in the vicinity of the BH spin axis, as we have mentioned above
\citep[see][]{2007ApJ...668..417F}. 
In the immediate vicinity of the BH $r\lesssim 5r_\mr{g}$, the tilt angle drastically decreases from \tr{$\sim 35^\circ$ to $\sim 15^\circ$}. 
This is due to the Frame-dragging effect, which acts to move the gas to the equatorial plane $(\theta=90^\circ)$. 
The precession angle becomes larger near the BH ($r\lesssim 15r_\mr{g}$).
This is probably because the LT effect is more pronounced. 

To summarize the structure of the accretion disk between \tr{$3.8\times 10^4 t_\mr{g}$ and $4.0\times10^4 t_\mr{g}$}, the accretion disk has the same tilt angle with the initial torus and precesses with \tg{$116^\circ$} in $r\gtrsim 15r_\mr{g}$. 
In $5r_\mr{g} \lesssim r \lesssim 15r_\mr{g}$, the closer to the BH, the larger the tilt and precession angles. 
The precession angle increases further, but the tilt angle decreases in $r\lesssim 5r_\mr{g}$. 
Such structures obtained in our GR-RMHD simulation is similar to those in \citep{2007ApJ...668..417F}. 

Figure \ref{t-tilt,prec} shows the time evolution of $\mathcal{T}$ and $\mathcal{P}$ at $r=30r_\mr{g}$ (solid line), $r=10r_\mr{g}$ (dashed line), and $r=5r_\mr{g}$ (dotted line). 
The results at $10r_\mr{g}$ and $5r_\mr{g}$ are plotted after \tr{$(t = 10^4 t_\mr{g})$}. 
In Figure \ref{t-tilt,prec}a, the tilt angle at $30r_\mr{g}$ is \tr{slightly decreases to $25^\circ$}. 
The tilt angles at $5r_\mr{g}$ and $10r_\mr{g}$ \tr{also slightly decrease and fluctuate.}
In addition, the figure shows that the smaller $r$ is, the larger $\cal{T}$ is.
This is because $\cal{T}$ becomes larger as it approaches the BH except in $r\lesssim 5r_\mr{g}$, as shown in Figure \ref{r-tilt,prec}a. 

In Figure \ref{t-tilt,prec}b, the precession angle at $r=30r_\mr{g}$ (solid line) initially matches that of the initial torus $(\mathcal{P}\sim \mathcal{P}_{0} = 180^\circ)$ and increases with time. 
At $r=10r_\mr{g} (5r_\mr{g})$, the precession angle also increases, keeping the difference of about $15^\circ (30^\circ)$ relative to the value at $r=30r_\mr{g}$. 
This is due to the LT effect.
Since the LT effect is more pronounced 
in the vicinity of the BH, 
the smaller the radius, the larger the precession angle (see Figure \ref{r-tilt,prec}b).

We can estimate the precession period at \tg{$1.2\times 10^5 t_\mr{g}$} since the accretion disk precesses with about \tg{$115^\circ$ in $4.0\times10^4 t_\mr{g}$}.
This period is consistent with the prediction
by the theory of the LT precession $(T_\mr{LT}=\pi/ar^3)\sim1.3\times 10^5t_\mr{g}$ at $r=33r_\mr{g}$, which is the radius of the maximum density of the initial torus. 

\subsection{\tr{Jet} and Radiation Flux}
Figures \ref{ph-th-flux}a and \ref{ph-th-flux}b 
show the $\phi-\theta$ profiles of kinetic energy flux density $(l_\mr{kin}c)$ in the \tr{jet} region at \tr{$r=800r_\mr{g}$ in $t= 10^4 t_\mr{g}$ and $t=4.0\times 10^4 t_\mr{g}$}, respectively.  
Here $l_\mr{kin}$ is the kinetic energy density defined as 
\begin{equation}
    l_\mr{kin} = \rho u^r\left( u_t+\sqrt{-g_{tt}} \right). 
\end{equation}
The kinetic energy flux density is high around $\theta \sim 30^\circ$ and \tr{$\phi\sim180^\circ$} as well as $\theta \sim 150^\circ$
and \tr{$\phi\sim0^\circ$}. 
The maximum value is about \tr{$10^{22.5}\ \mr{erg\ cm^{-2}\  s^{-1}}$ at $t=10^4t_\mr{g}$ and $10^{21.5}\ \mr{erg\ cm^{-2}\  s^{-1}}$ at $t=4.0\times10^4t_\mr{g}$}. 
It is found that the high kinetic energy flux density region moves to the right from
\tr{$10^4 t_\mr{g}$ to $4.0\times 10^4 t_\mr{g}$}. 
This precession of the \tr{jet}
caused by the precession of the accretion disks
(see below for details). 
Figures \ref{t-outflow}c and \ref{t-outflow}d show the $\phi-\theta$ profiles of the isotropic luminosity $(L_\mr{iso})$ at \tr{$r=800r_\mr{g}$} \tr{in $t= 10^4 t_\mr{g}$ and $4.0 \times 10^4 t_\mr{g}$}, respectively. 
The isotropic luminosity is defined as 
\begin{equation}
    L_\mr{iso}=4\pi r^2 \max(-{R^r}_t,0), 
\end{equation}
where ${R^r}_t$ is the radiation flux density. 
The region of high isotropic luminosity is almost the same as that of the high kinetic energy flux density. 
It moves to the right, which is similar to the kinetic energy flux density. 
Here we note that the maximum value isotropic luminosity is about \tr{$96L_\mr{Edd}$} and is \tr{60} times higher than the bolometric luminosity \tr{$(L_\mr{rad}\sim 1.7L_\mr{Edd})$}. 

Figure \ref{t-outflow} shows the polar angle $(\langle\mathcal{T}\rangle_\mr{kin})$ and azimuthal angle $(\langle\mathcal{P}\rangle_\mr{kin})$ of the propagation direction of the \tr{jet averaged over $r=500r_\mr{g}-800r_\mr{g}$}. 
Similarly, we plot the polar angle $(\langle\mathcal{T}\rangle_\mr{rad})$ and azimuthal angle $(\langle\mathcal{P}\rangle_\mr{rad})$ of the propagation direction of the radiation \tr{and the polar angle $(\langle\mathcal{T}\rangle_\mr{mag})$ and azimuthal angle $(\langle\mathcal{P}\rangle_\mr{mag})$ of the propagation direction of the magnetic flux}. 
Here, we evaluate $\langle\mathcal{T}\rangle_\mr{kin}$, $\langle\mathcal{P}\rangle_\mr{kin}$, $\langle\mathcal{T}\rangle_\mr{rad}$, $\langle\mathcal{P}\rangle_\mr{rad}$,
\tr{$\langle\mathcal{T}\rangle_\mr{mag}$, and $\langle\mathcal{P}\rangle_\mr{mag}$}
as 
\begin{equation}
    \cos\langle\mathcal{T}\rangle_{\rm kin}=
    \frac{\int^{2\pi}_{0}\int^{\pi/2}_{0} l_\mr{kin} \cos\theta  \sqrt{-g} d\theta d\phi } {\int^{2\pi}_{0}\int^{\pi/2}_{0} l_\mr{kin} \sqrt{-g} d\theta d\phi},
    \label{eq.th_kin}
\end{equation}
\begin{equation}
    \tan\langle\mathcal{P}\rangle_{\rm kin}=
    \frac{\int^{2\pi}_{0}\int^{\pi/2}_{0} l_\mr{kin} \sin\theta \cos\phi \sqrt{-g} d\theta d\phi } {\int^{2\pi}_{0}\int^{\pi/2}_{0} l_\mr{kin} \sin\theta \sin\phi \sqrt{-g} d\theta d\phi},
    \label{eq.phi_kin}
\end{equation}
\begin{equation}
    \cos\langle\mathcal{T}\rangle_{\rm rad}=
    \frac{\int^{2\pi}_{0}\int^{\pi/2}_{0} {R^r}_t \cos\theta \sqrt{-g} d\theta d\phi } {\int^{2\pi}_{0}\int^{\pi/2}_{0} {R^r}_t \sqrt{-g} d\theta d\phi},
    \label{eq.th_rad}
\end{equation}
\begin{equation}
    \tan\langle\mathcal{P}\rangle_{\rm rad}=
    \frac{\int^{2\pi}_{0}\int^{\pi/2}_{0} {R^r}_t \sin\theta \cos\phi \sqrt{-g} d\theta d\phi } {\int^{2\pi}_{0}\int^{\pi/2}_{0} {R^r}_t \sin\theta \sin\phi \sqrt{-g} d\theta d\phi},
    \label{eq.phi_rad}
\end{equation}
\tr{
\begin{equation}
    \cos\langle\mathcal{T}\rangle_{\rm mag}=
    \frac{\int^{2\pi}_{0}\int^{\pi/2}_{0} (b^2u^ru_t-b^rb_t) \cos\theta\sqrt{-g} d\theta d\phi } {\int^{2\pi}_{0}\int^{\pi/2}_{0} (b^2u^ru_t-b^rb_t) \sqrt{-g} d\theta d\phi},
    \label{eq.th_rad}
\end{equation}
and 
\begin{equation}
    \tan\langle\mathcal{P}\rangle_{\rm mag}=
    \frac{\int^{2\pi}_{0}\int^{\pi/2}_{0} (b^2u^ru_t-b^rb_t) \sin\theta \cos\phi \sqrt{-g} d\theta d\phi } {\int^{2\pi}_{0}\int^{\pi/2}_{0} (b^2u^ru_t-b^rb_t) \sin\theta \sin\phi \sqrt{-g} d\theta d\phi},
    \label{eq.phi_rad}
\end{equation}
} respectively. 
The integration of Equations (\ref{eq.th_kin})-(\ref{eq.phi_rad}) is performed 
only in the northern \tr{jet} region
($Be\geq0.05$ in $0^\circ\leq\theta\leq90^\circ$).
\tr{We plot tilt angle $(\langle\mathcal{T}\rangle_r)$ and precession angle $(\langle\mathcal{P}\rangle_r)$ of the disk averaged over $r=50r_\mr{g}-300r_\mr{g}$ as black lines.}

In Figure \ref{t-outflow}a, $\langle\mathcal{T}\rangle_\mr{rad}$, $\langle\mathcal{T}\rangle_\mr{kin}$, and
\tr{$\langle\mathcal{T}\rangle_\mr{mag}$  
slightly decrease with a decrease of the tilt angle and have a larger fluctuation than the tilt angle.}
 This angle is closer to the rotation axis \tr{$\mathcal{T}$} than the spin axis of the BH $(\theta=0^\circ)$. 
In Figure \ref{t-outflow}b, $\langle\mathcal{P}\rangle_\mr{rad}$, $\langle\mathcal{P}\rangle_\mr{kin}$, and \tr{$\langle\mathcal{P}\rangle_\mr{mag}$} increase with time on average with fluctuations of $\pm5^\circ$. 
The change of $\langle\mathcal{P}\rangle_\mr{rad}$ and $\langle\mathcal{P}\rangle_\mr{kin}$
obtained by the least-squares method is approximately \tr{$40^\circ$ and $56^\circ$
for a period between $t=10^4 t_\mr{g}$ and $4.0\times10^4 t_\mr{g}$, respectively. 
This amount of change is slightly smaller than similar to \tr{$\mathcal{P}$ $(\sim 63^\circ)$}. 
The change of $\mathcal{P}$ is smaller than that of the precession angle in Figure \ref{r-tilt,prec}. 
The smaller precession angle is due to averaging to the outer radius of the disk. }
\tr{The increasing of $\langle\mathcal{P}\rangle_\mr{kin}$ means the precessing jet forms with the precession period of $1.9\times 10^5 t_\mr{g}$.} 
From Figures \ref{ph-th-flux}d and \ref{t-outflow},
it is expected that the luminosity observed from $\theta \sim 20^\circ$ drastically varies with the period of about \tr{$2.7\times 10^5 t_\mr{g}=13.6\ \mr{s}\ (74\ \mr{mHz})$}, assuming the black hole mass of $10M_\odot$. 
We need long-term simulations to confirm this time variation, although the simulation time is much shorter than this period in the present paper. 

%% file: summary_v1.tex
\section{Summary and Discussion}
We perform the 3D GR-RMHD simulation of a super-Eddington accretion disk tilted with respect to the spin axis of the BH. 
As a result, the non-axisymmetric distorted accretion disk with a mass accretion rate of about \tr{$300L_\mr{Edd}$} forms. 
The accretion disk has the same tilt angle ($\sim 30^\circ$) as the initial torus in $r\gtrsim 15r_\mr{g}$. 
It is found that the tilt angle (angle between the disk rotation axis and the BH spin axis) is larger as closer to the BH, except in the very vicinity of the BH ($r\lesssim 5r_\mr{g}$), since the disk matter tends to accrete from the high latitude where the ISCO radius is larger than that at around the equatorial plane. 
\tr{In the very vicinity of the BH,}
the tilt angle is smaller due to the frame-dragging effect.
It is found that the precession angle (azimuthal angle between the disk rotation axis and the rotation axis of the initial torus) is larger as closer to the BH. 
For instance, \tr{it is about $20^\circ-30^\circ$ larger at around $r \sim 5r_\mr{g}$ than at around $r \sim 30r_\mr{g}$.}
This is caused by the Lense-Thirring effect becoming more effective the closer to the BH. 
The accretion disk is thought to precesses 
with a period of \tg{$\sim 1.2\times 10^5 t_\mr{g}$} keeping the above-distorted structure. 
In addition, the ejection direction of the \tr{jet},
launched from the super-Eddington accretion disk 
with the velocity of $0.3c$,
is closer to the rotation axis of the accretion disk in $r \gtrsim 15r_\mr{g}$ than the BH spin axis. 
This ejection direction precesses due to the precession of the disk. 
Radiation energy is also mainly released in approximately the same direction as \tr{the jet}.
This direction also changes with the ejection direction of the \tr{jet} via the precession of the disk. 
This means that the observed luminosity also oscillates
quasi-periodically.
\tr{The period is $\sim 2.7\times 10^5 t_\mr{g}$ estimated from the change of the azimuthal angle of the radiation propagation direction. 
} 

For the case of the stellar-mass black holes with $10M_\odot$,
the oscillation period of the isotropic luminosity expected by our simulation is about \tr{$13.6\ \mr{s}$ ($\sim 74$ mHz)} and roughly consistent with the quasi-periodic oscillations observed in some ULXs,
$10-40\ \mr{mHz}$ in NGC5408 X-1 and NGC6946 X-1, 
$80-630\ \mr{mHz}$ in NGC1313 X-1, 
and $\sim 650\ \mr{mHz}$ in IC342 X-1
\citep{2019MNRAS.486.2766A}. 
However, the time variation of the propagation direction 
of the \tr{jet} disagrees with the observations of V404 Cygni. 
By \cite{2019Natur.569..374M},
the time scale is about some minutes or hours, much longer than the precession period obtained from our simulation.
\tr{Since the time scale of the Lense-Thirring precession increases with an increase of the disk size, the observation might be explained if the more extended disk makes the precessional motion.
Conversely, the tidal disruption event, OJ 287, can be explained by the precession of the inner part of the disk \citep{1996A&A...315L..13S}.
The timescale for the luminosity variation of this object is 10yr, which is one-fortieth of our result, $\sim 400$yr, evaluated by assuming $10^6 M_\odot$ BH.
Therefore, if a disk with a size of $\sim 40 r_{\rm g}$ which is about half of our simulation precesses, the simulation result becomes consistent with the observations.
However, another tidal disruption event, Swift J1644, has an extremely short timescale and may be difficult to understand via the precession motion
\citep{2011Sci...333..203B, 2012Sci...337..949R}.
In this case, high-frequency QPO would be the likely mechanism.
}

\tr{
Here, we discuss the similarities and differences between the results of the precession of super-Eddington disks simulated by GR-RMHD in this study and that of less luminous disks investigated by GR-MHD simulations \citep{2005ApJ...623..347F, 2007ApJ...668..417F, 2018MNRAS.474L..81L}.
In both cases, the tilted and twisted disk appears, causing precessional motion.
Also, the precessional motion of the jet appears in both disks, but the acceleration mechanism of the jet is not the same.
Although the Blandford-Znajek mechanism induces the jet in \cite{2018MNRAS.474L..81L}, the jet is mainly powered by the radiation force in the present study. 
We note that if the magnetic flux at the event horizon increases in further long-term simulation, the Blandford-Znajek process may contribute significantly to jet formation.
The precession of the collimated radiation flux is considered to be a specific feature of a geometrically and optically thick super-Eddington disk. If the disk were optically thin like a less luminous disk, the radiation collimation would not be so pronounced.
Our precession period of the disk in $15r_\mr{g}<r<40r_\mr{g}$, $1.2 t_\mr{g} \sim 6\rm{s} \left(M/10M_\odot\right)$, is comparable to that in \cite{2007ApJ...668..417F}, $\sim 3\ \mr{s}$, but is 10 times shorter than that in \cite{2018MNRAS.474L..81L}.
One of the reasons for the difference in period would be the difference in initial conditions (e.g., torus size).
A large disk formed from a large torus should exhibit a long precession period. 
However, the gradual decrease of the precession rate, which has been reported by \cite{2018MNRAS.474L..81L} but does not appear in our study, cannot be explained solely by differences in initial conditions. We will discuss below.
}

\tr{
The discrepancy in the precession rate, which is nearly constant in our simulation and gradually decreases in \cite{2018MNRAS.474L..81L}, may be due to the decrease in the disk viscosity.
The sound speed of the hot and less luminous accretion flow such as the radiatively inefficient accretion flow (RIAF), 
which can be studied by GR-MHD simulations, is about $0.3c$ since the proton temperature is around $10^{12}$K.
In the present simulation, the radiative cooling reduces the sound speed to $0.05c$ so the viscous timescale is estimated to be 6 times longer than that of the GR-MHD simulations.
The small viscosity may suppress the disk extension and keep the precession rate constant for a long time.
However, we should note that our simulation may underestimate the extent of the disk due to insufficient resolution.
In our simulation, the quality factor in $\theta$-direction is about 10 around the inner radius of the initial torus.
Also, this factor is $\sim 10$ in $\rho>10^{-4}\rho_0$ region and $\sim 5$ in $\rho>10^{-3}\rho_0$ region at $t \sim 4.0 \times 10^4t_\mr{g}$.
On the other hand, the factor in $\phi$-direction for that time is about half of those in $\theta$-direction.
This means that the resolution of the present simulation is not extremely low, but it is not sufficient to treat the MRI accurately.
Therefore, although the disk continues extending
(the density-weighted average radius increases from $49 r_\mr{g}$ to $94 r_\mr{g}$ between $t=0$ and $4\times 10^4 t_\mr{g}$),
the underestimation of the magnetic viscosity may induce the underestimation of the rate of disk size growth.
This might keep the precession rate constant without decreasing.
Here we note that the amount of gas that flowed out of the outer boundary is only about 1\% of the initial torus and that the size of the disk is not artificially reduced via the mass ejection.
We need high-resolution simulations to accurately treat MRI and solve this problem.
We also stress that simulations under realistic conditions would be necessary since the actual disk is not extended by the torus but is formed by the gas supplied from the outer region.
}

\tr{
Since the shape and precession timescale of the tilted super-Eddington disk are thought to depend on the initial torus setting as position, density, magnetic field, and tilt angle,
the initial torus dependence should be investigated in further simulations.
In addition, we should perform long-term simulations 
that exceed the precession timescale and confirm whether the \tr{jet} and luminosity oscillate periodically. 
Improving the solution method for radiative transfer is also important. 
The M1-closure method employed in the present study is known not to give correct solutions in optically thin and extremely anisotropic radiation field situations \citep{2020ApJ...901...96A} since the closure relation is simply calculated without using specific radiation intensity.
In the case of super-Eddington flows, radiation fields in the funnel region may be inaccurate.
This problem can be fixed by directly solving the radiation transfer equation to obtain the specific intensity
\citep{1992ApJS...80..819S, 2014ApJS..213....7J, 2016ApJ...818..162O, 2020ApJ...901...96A}. 
The GR-RMHD simulations coupling such radiative transfer method with GR-MHD have already been performed by \cite{2022ApJ...929...93A} and \cite{2023ApJ...949..103W}.}

In the present work, we mainly evaluate the kinetic luminosity and the photon luminosity at $r=800r_{\rm g}$ which are not sensitive to the radius in such a distant region. 
However, the radiation force may accelerate the gas gradually over a long distance. 
This is the case, the photon luminosity would decrease and conversely kinetic luminosity would increase. 
In order to more accurately determine the energy released from the system, simulations with larger computational domain should be performed. 
Such simulations are also useful in making direct comparisons with observations. 
In the present work, the isotropic luminosity is used as a measure of the anisotropy of the radiation but is different from the luminosity detected by a distant observer.
\trb{In addition, the position of the photosphere estimated from the radially extrapolated density is roughly $10^3r_\mathrm{g}$, which is close to the outer boundary of the present simulations. 
}
To calculate the observed photon luminosity, it is necessary to perform simulations with the large domain that includes the photosphere and to perform radiation transfer calculations to obtain the specific intensity. Such simulations are left as important future work.

\trb{High-resolution simulations are also an important future work. In the present study, we use a first-order upwind method for the numerical fluxes at the event horizon and a first-order Lax-Friedrichs method for the three meshes outside it. This method has the advantage of not needing to refer to information inside the event horizon, but the numerical diffusion may affect the results. This problem could be resolved with high-resolution simulations that provide a sufficient number of small cells near the horizon. Such simulations remain a future work.}

Our simulations are conducted with Cray XC50 at the Center for Computational Astrophysics (CfCA), National Astronomical Observatory of Japan (NAOJ), Oakforest-PACS at the CCS, University of Tsukuba, and with Wisteria/BDEC-01 Odyssey (the University of Tokyo), provided by the Multidisciplinary Cooperative Research Program in the Center for Computational Sciences, University of Tsukuba.. 
This work was supported by JSPS KAKENHI Grant Numbers 23K03445(Y.A.), 21H01132(R.T.), 21H04488, 18K03710(K.O.). 
\tr{This work was also supported by MEXT as “Program for Promoting Researches on the Supercomputer Fugaku” (Structure and Evolution of the Universe Unraveled by Fusion of Simulation and AI; Grant Number JPMXP1020230406) and used computational resources of supercomputer Fugaku provided by the RIKEN Center for Computational Science (Project ID:hp230204, hp230116).}

%% file: figures_v1.tex
\begin{figure}[!ht]
\centering
\includegraphics[scale=0.6]{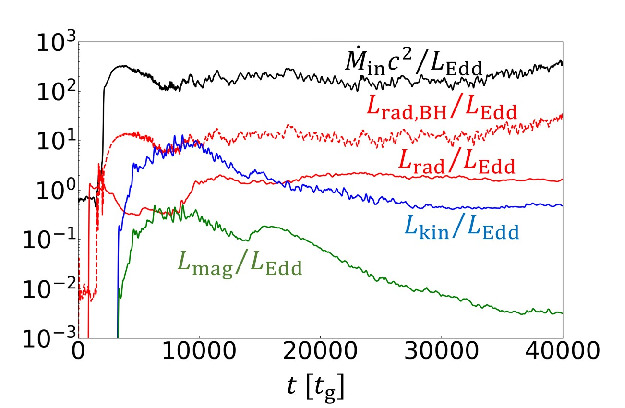}
\caption{Time evolution of the mass accretion rate (black), kinetic luminosity (blue), \tr{electromagnetic luminosity (green),} photon luminosity (solid red), and photon luminosity swallowed by the BH (dashed red). }
\label{t-mdot}
\end{figure}

\begin{figure}[!ht]
\centering
\includegraphics[scale=0.6]{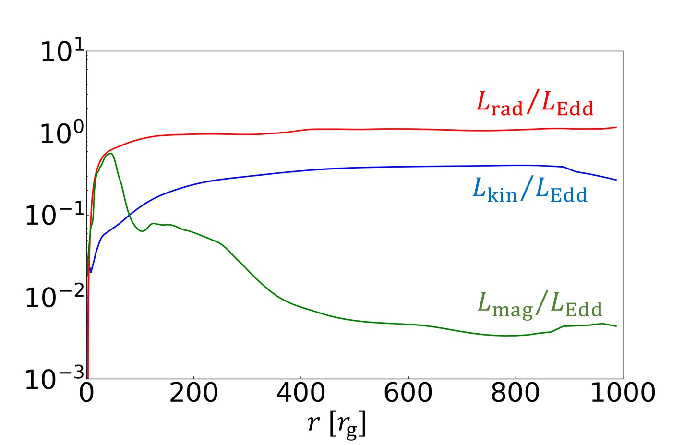}
\caption{\tra{Radial profile of the time-averaged photon luminosity (red), kinetic luminosity (blue), and electromagnetic luminosity (green) between $3.8\times10^4t_\mr{g}$ and $4.0\times10^4t_\mr{g}$ }}
\label{r-lum}
\end{figure}

\begin{figure}[!ht]
\centering
\includegraphics[scale=0.6]{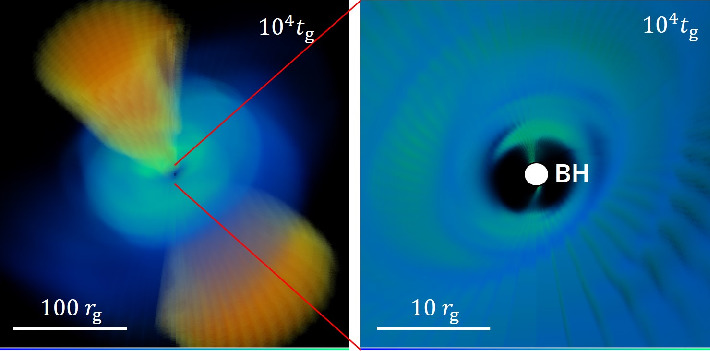}
\caption{Left panel shows volume rendered density (blue) and Lorentz factor (orange). Right panel shows the density distribution near the BH (white filled circle).}
\label{volume-rendering}
\end{figure}

\begin{figure}[!ht]
\centering
\includegraphics[scale=0.6]{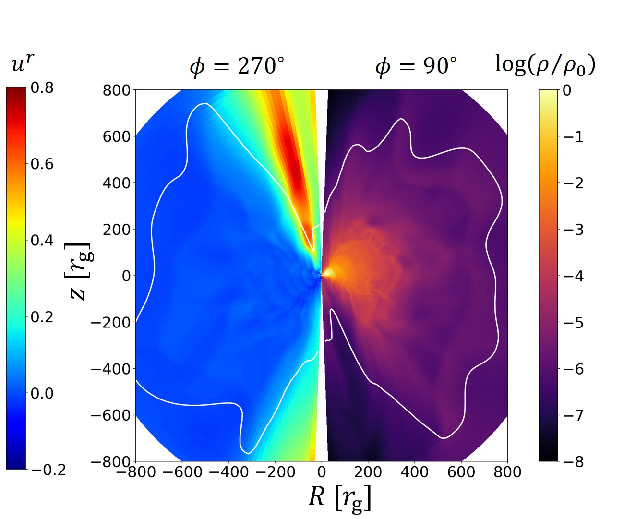}
\caption{\tra{Profiles of radial velocity (left) and density (right) at $t=4.0\times 10^4 t_\mr{g}$. White contour shows the photosphere measured from the outer boundary.}
}
\label{ur-ro}
\end{figure}

\begin{figure}[!ht]
\centering
\includegraphics[scale=0.6]{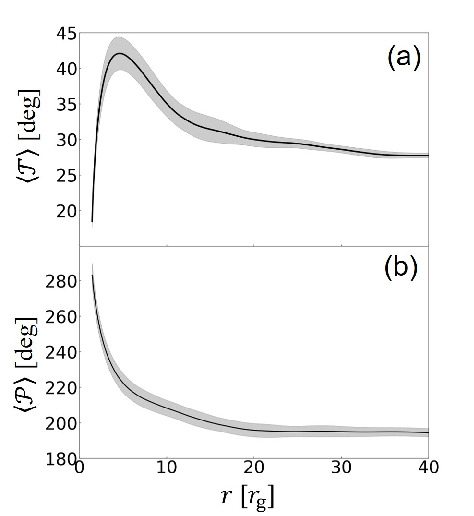}
\caption{Radial profile of (a) the time-averaged tilted angle $\langle \mathcal{T} \rangle$ and (b) time-averaged precession angle $\langle \mathcal{P} \rangle$. The shaded region indicates $1\sigma$-variation. 
}
\label{r-tilt,prec}
\end{figure}

\begin{figure}[!ht]
\centering
\includegraphics[scale=0.6]{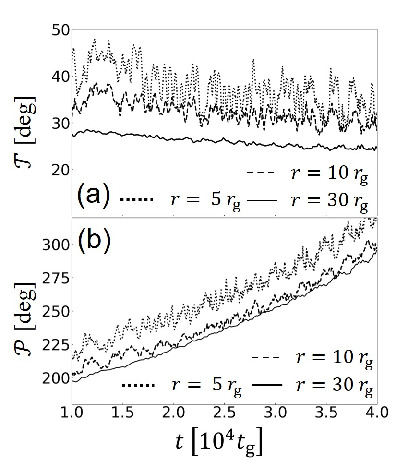}
\caption{Time evolution of (a) the tilt angle and (b) precession angle at $r=30r_\mr{g}$ (solid), $r=10r_\mr{g}$ (dashed), and $r=5r_\mr{g}$ (dotted). }
\label{t-tilt,prec}
\end{figure}

\begin{figure}[!ht]
\centering
\includegraphics[scale=0.55]{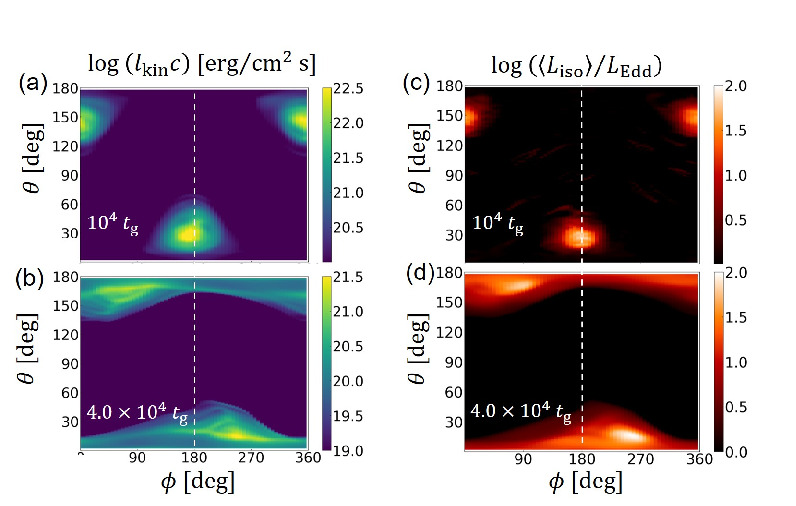}
\caption{Profiles of the kinetic energy density flux of the outflow at (a) \tr{$t= 10^4t_\mr{g}$ and (b) $t=4.0\times 10^4t_\mr{g}$} in $\phi-\theta$ plane.
Profiles of the time-averaged isotropic luminosity at (c) \tr{$t=10^4 t_\mr{g}$ and (d) $t=4.0\times 10^4 t_\mr{g}$. These are measured at $r=800r_\mr{g}$}}
\label{ph-th-flux}
\end{figure}

\begin{figure}[!ht]
\centering
\includegraphics[scale=0.6]{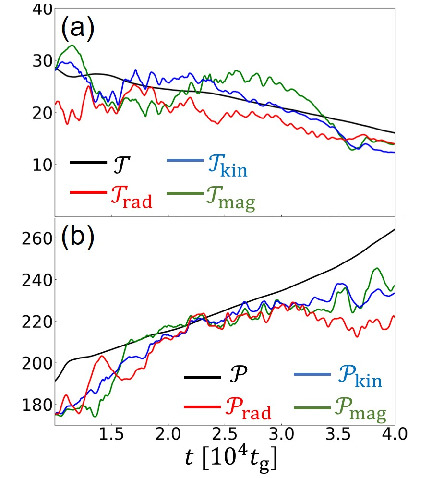}
\caption{
(a) Time evolution of the polar angle of the propagation direction of the outflow (blue), \tr{that of the radiation (red), and that of the magnetic flux (green) averaged over $r=500r_\mr{g} - 800r_\mr{g}$. 
The black line represents the tilt angle of the accretion disk \tr{averaged over $r=50r_\mr{g} - 300r_\mr{g}$}}. 
(b) Time evolution of the azimuthal angle of the propagation direction of the outflow (blue), \tr{that of the radiation (red), and that of the magnetic flux (green) averaged over $r=500r_\mr{g} - 800r_\mr{g}$}.
The black line indicates the precession angle of the disk \tr{averaged over $r=50r_\mr{g} - 300r_\mr{g}$}. }
\label{t-outflow}
\end{figure}

%% file: asahina_v1.bbl
\begin{thebibliography}{42}
\expandafter\ifx\csname natexlab\endcsname\relax\def\natexlab#1{#1}\fi

\bibitem[{{Armitage} \& {Natarajan}(1999)}]{1999ApJ...525..909A}
{Armitage}, P.~J., \& {Natarajan}, P. 1999, \apj, 525, 909

\bibitem[{{Asahina} \& {Ohsuga}(2022)}]{2022ApJ...929...93A}
{Asahina}, Y., \& {Ohsuga}, K. 2022, \apj, 929, 93

\bibitem[{{Asahina} {et~al.}(2020){Asahina}, {Takahashi}, \& {Ohsuga}}]{2020ApJ...901...96A}
{Asahina}, Y., {Takahashi}, H.~R., \& {Ohsuga}, K. 2020, \apj, 901, 96

\bibitem[{{Atapin} {et~al.}(2019){Atapin}, {Fabrika}, \& {Caballero-Garc{\'\i}a}}]{2019MNRAS.486.2766A}
{Atapin}, K., {Fabrika}, S., \& {Caballero-Garc{\'\i}a}, M.~D. 2019, \mnras, 486, 2766

\bibitem[{{Bardeen} \& {Petterson}(1975)}]{1975ApJ...195L..65B}
{Bardeen}, J.~M., \& {Petterson}, J.~A. 1975, \apjl, 195, L65

\bibitem[{{Blandford} \& {Znajek}(1977)}]{1977MNRAS.179..433B}
{Blandford}, R.~D., \& {Znajek}, R.~L. 1977, \mnras, 179, 433

\bibitem[{{Bloom} {et~al.}(2011){Bloom}, {Giannios}, {Metzger}, {Cenko}, {Perley}, {Butler}, {Tanvir}, {Levan}, {O'Brien}, {Strubbe}, {De Colle}, {Ramirez-Ruiz}, {Lee}, {Nayakshin}, {Quataert}, {King}, {Cucchiara}, {Guillochon}, {Bower}, {Fruchter}, {Morgan}, \& {van der Horst}}]{2011Sci...333..203B}
{Bloom}, J.~S., {Giannios}, D., {Metzger}, B.~D., {et~al.} 2011, Science, 333, 203

\bibitem[{{Eggum} {et~al.}(1988){Eggum}, {Coroniti}, \& {Katz}}]{1988ApJ...330..142E}
{Eggum}, G.~E., {Coroniti}, F.~V., \& {Katz}, J.~I. 1988, \apj, 330, 142

\bibitem[{{Fishbone} \& {Moncrief}(1976)}]{1976ApJ...207..962F}
{Fishbone}, L.~G., \& {Moncrief}, V. 1976, \apj, 207, 962

\bibitem[{{Fragile} \& {Anninos}(2005)}]{2005ApJ...623..347F}
{Fragile}, P.~C., \& {Anninos}, P. 2005, \apj, 623, 347

\bibitem[{{Fragile} {et~al.}(2007){Fragile}, {Blaes}, {Anninos}, \& {Salmonson}}]{2007ApJ...668..417F}
{Fragile}, P.~C., {Blaes}, O.~M., {Anninos}, P., \& {Salmonson}, J.~D. 2007, \apj, 668, 417

\bibitem[{{Gonz{\'a}lez} {et~al.}(2007){Gonz{\'a}lez}, {Audit}, \& {Huynh}}]{2007A&A...464..429G}
{Gonz{\'a}lez}, M., {Audit}, E., \& {Huynh}, P. 2007, \aap, 464, 429

\bibitem[{{Hawley} \& {Krolik}(2002)}]{2002ApJ...566..164H}
{Hawley}, J.~F., \& {Krolik}, J.~H. 2002, \apj, 566, 164

\bibitem[{{Jiang} {et~al.}(2014){Jiang}, {Stone}, \& {Davis}}]{2014ApJS..213....7J}
{Jiang}, Y.-F., {Stone}, J.~M., \& {Davis}, S.~W. 2014, \apjs, 213, 7

\bibitem[{{Kobayashi} {et~al.}(2018){Kobayashi}, {Ohsuga}, {Takahashi}, {Kawashima}, {Asahina}, {Takeuchi}, \& {Mineshige}}]{2018PASJ...70...22K}
{Kobayashi}, H., {Ohsuga}, K., {Takahashi}, H.~R., {et~al.} 2018, \pasj, 70, 22

\bibitem[{{Laing} {et~al.}(2008){Laing}, {Bridle}, {Parma}, {Feretti}, {Giovannini}, {Murgia}, \& {Perley}}]{2008MNRAS.386..657L}
{Laing}, R.~A., {Bridle}, A.~H., {Parma}, P., {et~al.} 2008, \mnras, 386, 657

\bibitem[{{Liska} {et~al.}(2018){Liska}, {Hesp}, {Tchekhovskoy}, {Ingram}, {van der Klis}, \& {Markoff}}]{2018MNRAS.474L..81L}
{Liska}, M., {Hesp}, C., {Tchekhovskoy}, A., {et~al.} 2018, \mnras, 474, L81

\bibitem[{{Liska} {et~al.}(2021){Liska}, {Hesp}, {Tchekhovskoy}, {Ingram}, {van der Klis}, {Markoff}, \& {Van Moer}}]{2021MNRAS.507..983L}
---. 2021, \mnras, 507, 983

\bibitem[{{Liska} {et~al.}(2019){Liska}, {Tchekhovskoy}, {Ingram}, \& {van der Klis}}]{2019MNRAS.487..550L}
{Liska}, M., {Tchekhovskoy}, A., {Ingram}, A., \& {van der Klis}, M. 2019, \mnras, 487, 550

\bibitem[{{Liska} {et~al.}(2023){Liska}, {Kaaz}, {Musoke}, {Tchekhovskoy}, \& {Porth}}]{2023ApJ...944L..48L}
{Liska}, M.~T.~P., {Kaaz}, N., {Musoke}, G., {Tchekhovskoy}, A., \& {Porth}, O. 2023, \apjl, 944, L48

\bibitem[{{Machida} \& {Matsumoto}(2008)}]{2008PASJ...60..613M}
{Machida}, M., \& {Matsumoto}, R. 2008, \pasj, 60, 613

\bibitem[{{McKinney} {et~al.}(2014){McKinney}, {Tchekhovskoy}, {Sadowski}, \& {Narayan}}]{2014MNRAS.441.3177M}
{McKinney}, J.~C., {Tchekhovskoy}, A., {Sadowski}, A., \& {Narayan}, R. 2014, \mnras, 441, 3177

\bibitem[{{Middleton} {et~al.}(2011){Middleton}, {Roberts}, {Done}, \& {Jackson}}]{2011MNRAS.411..644M}
{Middleton}, M.~J., {Roberts}, T.~P., {Done}, C., \& {Jackson}, F.~E. 2011, \mnras, 411, 644

\bibitem[{{Miller-Jones} {et~al.}(2019){Miller-Jones}, {Tetarenko}, {Sivakoff}, {Middleton}, {Altamirano}, {Anderson}, {Belloni}, {Fender}, {Jonker}, {K{\"o}rding}, {Krimm}, {Maitra}, {Markoff}, {Migliari}, {Mooley}, {Rupen}, {Russell}, {Russell}, {Sarazin}, {Soria}, \& {Tudose}}]{2019Natur.569..374M}
{Miller-Jones}, J. C.~A., {Tetarenko}, A.~J., {Sivakoff}, G.~R., {et~al.} 2019, \nat, 569, 374

\bibitem[{{Musoke} {et~al.}(2023){Musoke}, {Liska}, {Porth}, {van der Klis}, \& {Ingram}}]{2023MNRAS.518.1656M}
{Musoke}, G., {Liska}, M., {Porth}, O., {van der Klis}, M., \& {Ingram}, A. 2023, \mnras, 518, 1656

\bibitem[{{Ohsuga} \& {Mineshige}(2011)}]{2011ApJ...736....2O}
{Ohsuga}, K., \& {Mineshige}, S. 2011, \apj, 736, 2

\bibitem[{{Ohsuga} {et~al.}(2009){Ohsuga}, {Mineshige}, {Mori}, \& {Kato}}]{2009PASJ...61L...7O}
{Ohsuga}, K., {Mineshige}, S., {Mori}, M., \& {Kato}, Y. 2009, \pasj, 61, L7

\bibitem[{{Ohsuga} {et~al.}(2005){Ohsuga}, {Mori}, {Nakamoto}, \& {Mineshige}}]{2005ApJ...628..368O}
{Ohsuga}, K., {Mori}, M., {Nakamoto}, T., \& {Mineshige}, S. 2005, \apj, 628, 368

\bibitem[{{Ohsuga} \& {Takahashi}(2016)}]{2016ApJ...818..162O}
{Ohsuga}, K., \& {Takahashi}, H.~R. 2016, \apj, 818, 162

\bibitem[{{Okuda} {et~al.}(1997){Okuda}, {Fujita}, \& {Sakashita}}]{1997PASJ...49..679O}
{Okuda}, T., {Fujita}, M., \& {Sakashita}, S. 1997, \pasj, 49, 679

\bibitem[{{Reis} {et~al.}(2012){Reis}, {Miller}, {Reynolds}, {G{\"u}ltekin}, {Maitra}, {King}, \& {Strohmayer}}]{2012Sci...337..949R}
{Reis}, R.~C., {Miller}, J.~M., {Reynolds}, M.~T., {et~al.} 2012, Science, 337, 949

\bibitem[{{Sadowski} \& {Narayan}(2016)}]{2016MNRAS.456.3929S}
{Sadowski}, A., \& {Narayan}, R. 2016, \mnras, 456, 3929

\bibitem[{{Sadowski} {et~al.}(2014){Sadowski}, {Narayan}, {McKinney}, \& {Tchekhovskoy}}]{2014MNRAS.439..503S}
{Sadowski}, A., {Narayan}, R., {McKinney}, J.~C., \& {Tchekhovskoy}, A. 2014, \mnras, 439, 503

\bibitem[{{Sillanpaa} {et~al.}(1996){Sillanpaa}, {Takalo}, {Pursimo}, {Nilsson}, {Heinamaki}, {Katajainen}, {Pietila}, {Hanski}, {Rekola}, {Kidger}, {Boltwood}, {Turner}, {Robertson}, {Honeycut}, {Efimov}, {Shakhovskoy}, {Fiorucci}, {Tosti}, {Ghisellini}, {Raiteri}, {Villata}, {de Francesco}, {Lanteri}, {Chiaberge}, {Peila}, \& {Heidt}}]{1996A&A...315L..13S}
{Sillanpaa}, A., {Takalo}, L.~O., {Pursimo}, T., {et~al.} 1996, \aap, 315, L13

\bibitem[{{S{\k{a}}dowski} {et~al.}(2016){S{\k{a}}dowski}, {Lasota}, {Abramowicz}, \& {Narayan}}]{2016MNRAS.456.3915S}
{S{\k{a}}dowski}, A., {Lasota}, J.-P., {Abramowicz}, M.~A., \& {Narayan}, R. 2016, \mnras, 456, 3915

\bibitem[{{Stella} {et~al.}(1999){Stella}, {Vietri}, \& {Morsink}}]{1999ApJ...524L..63S}
{Stella}, L., {Vietri}, M., \& {Morsink}, S.~M. 1999, \apjl, 524, L63

\bibitem[{{Stone} {et~al.}(1992){Stone}, {Mihalas}, \& {Norman}}]{1992ApJS...80..819S}
{Stone}, J.~M., {Mihalas}, D., \& {Norman}, M.~L. 1992, \apjs, 80, 819

\bibitem[{{Takahashi} \& {Ohsuga}(2015)}]{2015PASJ...67...60T}
{Takahashi}, H.~R., \& {Ohsuga}, K. 2015, \pasj, 67, 60

\bibitem[{{Takahashi} {et~al.}(2016){Takahashi}, {Ohsuga}, {Kawashima}, \& {Sekiguchi}}]{2016ApJ...826...23T}
{Takahashi}, H.~R., {Ohsuga}, K., {Kawashima}, T., \& {Sekiguchi}, Y. 2016, \apj, 826, 23

\bibitem[{{Takeuchi} {et~al.}(2010){Takeuchi}, {Ohsuga}, \& {Mineshige}}]{2010PASJ...62L..43T}
{Takeuchi}, S., {Ohsuga}, K., \& {Mineshige}, S. 2010, \pasj, 62, L43

\bibitem[{{Takeuchi} {et~al.}(2013){Takeuchi}, {Ohsuga}, \& {Mineshige}}]{2013PASJ...65...88T}
---. 2013, \pasj, 65, 88

\bibitem[{{White} {et~al.}(2023){White}, {Mullen}, {Jiang}, {Davis}, {Stone}, {Morozova}, \& {Zhang}}]{2023ApJ...949..103W}
{White}, C.~J., {Mullen}, P.~D., {Jiang}, Y.-F., {et~al.} 2023, \apj, 949, 103

\end{thebibliography}
